\documentclass[10pt]{article}
\usepackage{graphicx}

\usepackage{color}

\usepackage{lineno}

\title{Stereochemical configuration and selective excitation of the chiral molecule halothane}

\author{Martin Pitzer$^{1,2}$ $^\ast$, 
Gregor Kastirke$^{1}$, 
Phillip Burzynski$^{1}$,\\
Miriam Weller$^{1}$,
Daniel Metz$^{1}$,
Jonathan Neff$^{1}$,\\
Markus Waitz$^{1}$,
Florian Trinter$^{1}$,
Lothar Ph. H. Schmidt$^{1}$,\\
Joshua B. Williams$^{3}$,
Till Jahnke$^{1}$,
Horst Schmidt-B{\"o}cking$^{1}$,\\
Robert Berger$^{4}$,
Reinhard D{\"o}rner$^{1}$,
Markus Sch{\"o}ffler$^{1}$\\
\normalsize{$^{1}$Institute for Nuclear Physics,}
\normalsize{Johann Wolfgang Goethe-University Frankfurt,} \\
\normalsize{Max-von-Laue-Stra{\ss}e 1, 60438 Frankfurt, Germany}\\
\normalsize{$^{2}$Experimental Physics IV,}
\normalsize{University of Kassel,}\\
\normalsize{Heinrich-Plett-Stra{\ss}e 40, 34132 Kassel, Germany}\\
\normalsize{$^{3}$Department of Physics,} 
\normalsize{University of Nevada, Reno,} \\
\normalsize{1664 N. Virginia Street, Reno, NV 89557, USA}\\
\normalsize{$^{4}$Fachbereich Chemie,} 
\normalsize{Philipps-Universit{\"a}t,} \\
\normalsize{Hans-Meerwein-Stra{\ss}e, 35032 Marburg, Germany}\\
\\
\normalsize{$^\ast$To whom correspondence should be addressed; }\\
\normalsize{E-mail:  pitzer@atom.uni-frankfurt.de,} \\
}

\date{\today}

\begin{document}

\newcommand{\mpitzer}[1]{\textcolor{blue}{\emph{#1}}}

\maketitle
%%%%%%%%%%%%%%%%%%%%%%
%     Abstract
%%%%%%%%%%%%%%%%%%%%%%

\begin{abstract}
X-ray single-photon ionization and fragmentation of the chiral molecule halothane (CHBrClCF${}_{3}$) from a racemic mixture have been investigated using the COLTRIMS (Cold Target Recoil Ion Momentum Spectroscopy) technique. Two important facets related to the core ionization of this species are examined: Firstly, the distinction of enantiomers (mirror isomers) and the determination of absolute configuration on a single-molecule level by four-body Coulomb explosion; secondly, the interplay of site-selective excitation and fragmentation patterns. These results are easily transferrable to other molecular species and show the wealth of features that can be investigated by coincidence spectroscopy of chiral molecules.
\end{abstract}
\textbf{Keywords:}
Coulomb Explosion Imaging, Cold Target Recoil Ion Momentum Spectroscopy (COLTRIMS), Chirality, Absolute Configuration, Photoionization, Synchrotron, Core Excitation \\

\section{Introduction}
Symmetries are one of the most intriguing phenomena for the human spirit, be it in arts, in music or as a principle in our description of nature. Especially in physics, it had been assumed since the days of Isaac Newton that space and time were symmetric, i.e. that a space-time inversion of a closed physical system would not change its intrinsic interactions and dynamics. It was the discovery of the parity violation, first theoretically by Lee and Yang~\cite{Lee1956} and immediately afterwards by the experiment of Wu and co-workers~\cite{Wu1957} that put an end to this presumed certainty.\\
From the point of view of a chemist, and especially a biochemist and molecular biologist, the preponderance of a certain spatial direction or configuration is a fact known since the middle of the 19th century. At that time, chiral molecules were discovered, i.e. molecules that occur in two mirror image structures. For most chiral species, only one of these so-called enantiomers was found in nature, a circumstance known as biological homochirality.\\
After the discovery of parity violation, hypotheses to link this fundamental asymmetry to biological homochirality quickly arose. These hypotheses can be categorized into two main lines of thought: On the one hand, parity violation introduces a small difference in the energy levels of the two enantiomers~\cite{Quack2002}. On the other hand, the asymmetry of the weak-interaction-induced $\beta$-decay could have favoured one enantiomer over the other (Vester-Ulbricht hypothesis)~\cite{Vester1959}.\\
So far, none of these explanations have been able to establich an unequivocal link between the fundamental parity violation and the biological homochirality on earth. Many calculations and great experimental efforts have been devoted to detect the parity violating energy difference $\Delta_{\mathrm{PV}}E$, but due to the extreme weakness of the effect (the expected relative shift of spectral lines $\Delta_{\mathrm{PV}}E/(h\nu)$ being below $10^{-14}$ for experimentally accessible species~\cite{Darquie2010}), no observation has been made yet. Some authors conclude that the parity violation of the weak interaction is entirely independent from the one observed on the biological level~\cite{Bonner2000}.\\
The investigation of chiral molecules is not only of fundamental but also of practical interest. It has been reported for various pharmaceutical substances that only one of the enantiomers is of therapeutical use (e.g. ethambutol~\cite{Wilkinson1961}) or that the wrong enantiomer is even toxic (e.g. penicillamine~\cite{Walshe1992}). Much effort has been devoted to developing enantiopure substances~\cite{McConnell2007} and a wealth of techniques have been developed to distinguish and separate the enantiomers of a given species (see. e.g.~\cite{Wolf2007}).\\
Even though these techniques are well established, they do not yield direct information on the handedness of the molecular structure under investigation because only a macroscopic effect is measured. Accordingly, a particularly intriguing question is left open: Which sign of the observed effect is correlated to which of the two possible microscopic structures? This issue is known as the problem of absolute configuration. Most current techniques that distinguish between enantiomers need additional information to infer the absolute configuration. A direct determination, without reliance on theoretical models or semi-empirical rules, has only been possible with anomalous X-ray diffraction as introduced by Bijvoet et al.~\cite{Bijvoet1951}. Whereas the sign of the macroscopic effect (e.g. the optical rotation) leads to designation by (+)/(-) or D/L, Cahn, Ingold and later Prelog coined their famous rules~\cite{Cahn1951} to determine if a molecular structure is right-handed $(R)$ or left-handed $(S)$.\\
Chiral molecules have increasingly attracted attention in the physics community in the last two decades because gas-phase techniques have been developed that allow to study isolated chiral molecules. Under these circumstances, solvent and collective effects can be neglected and a deeper understanding of the molecules themselves and their light-matter interaction is possible. A list of major techniques encompasses Photoelectron Circular Dichroism (PECD)~\cite{Janssen2014}, microwave spectroscopy~\cite{Patterson2013}, and several mass-spectroscopic methods~\cite{Horsch2011, Zehnacker2010}.\\
A new prospect for investigating the relation between structure and dynamics was opened recently when coincidence experiments successfully showed that the two enantiomers of simple chiral species could be distinguished on the level of individual molecules~\cite{Pitzer2013}. These measurements employed so-called Coulomb Explosion Imaging (CEI)~\cite{Vager1989} to gain information on the molecular configuration. When the multiple ionization of a molecule occurs on a timescale shorter than the nuclear motion, the mutual repulsion of the fragment ions can be approximated by the Coulomb interaction. This, in turn, implies that their momentum directions are correlated to the relative positions at the time of ionization. In the measurements mentioned above, a femtosecond laser pulse was employed to multiply ionize the chiral prototype CHBrClF; the momentum vectors of all five atomic cations were measured in coincidence. These momentum vectors yielded a clearly separated signal from the two enantiomers. An experiment using foil-induced Coulomb Explosion Imaging was almost simultaneously able to determine the absolute configuration of a deuterated chiral epoxide~\cite{Herwig2013}.\\
Similar coincidence techniques have been employed to determine the absolute configuration of laser-aligned molecules~\cite{Christensen2015} and to investigate photoelectron circular dichroism for specific ionic fragments~\cite{Fanood2015}. The coincident measurement of electron and ion momenta allows to define new observables that are sensitive to symmetry violations and to the role of electron dynamics for molecular properties~\cite{Trinter2012}.\\
Recently, it has been demonstrated that Coulomb explosion of a chiral species can efficiently be induced by core ionization with an X-ray photon from a synchrotron source~\cite{Pitzer2016}. There, the distinction of enantiomers is even superior to the results obtained with a femtosecond laser. This is attributed to the shorter time-scale of the multiple ionization: The electronic processes after core excitation (e.g. Auger cascades) are expected to be faster (a few fs) than the pulse duration of a femtosecond laser pulse (around 40~fs).\\
In this work we extend our method to a slightly more complex molecule, the chiral ethane derivative halothane CHBrClCF${}_{3}$. The absolute configuration is supposed to be $(S)$-(+)-halothane~\cite{Ramig2012}. World Health Organization's \textit{Model Lists of Essential Medicines}   includes halothane~\cite{WHO2015}, although its use is decreasing due to its hepatotoxicity. Concerning stereospecific effects, a study reports slightly different physiological effects of the two enantiomers~\cite{Harris1998}. Inner shell photoionization and subsequent fragmentation have been investigated by Souza et al.~\cite{Souza2001}. Halothane showcases two aspects that become relevant when extending Coulomb Explosion Imaging to larger molecules: Can the configuration be determined again on a single-molecule level for a molecule that consists not only of a single carbon center? How does the fragmentation dynamics change depending on which carbon atom is initially excited and can we use this to enhance relevant fragmentation channels for the determination of absolute configuration?\\
Concerning the first question, results from our previous experiments show that molecular fragments (in contrast to atomic ions) can also be used for the determination of absolute configuration~\cite{Pitzer2016}. A similar approach has been employed, for example, to extract the photodynamics of deuterated benzene from the momentum correlations of molecular fragments~\cite{Matsuda2009}.\\
The second aspect, the interplay between selective excitation and fragmentation pathways has been investigated since the early days of synchrotron radiation and electron-ion coincidence spectroscopy~\cite{Eberhardt1983}. Particularly interesting for the investigation of halothane is the observation by Habenicht et al.~\cite{Habenicht1991} that the yield of CF${}_{3}^{+}$ from trifluoroethane (CF${}_{3}^{+}$CH${}_{3}^{+}$) increases significantly for excitation of the 'opposite' carbon atom C${}_{\mathrm{H}}$. More recent experiments employ PEPIPICO-spectra (Photoelectron-Photoion-Photoion-Coincidence), e.g. to investigate the conditions for which site-specific fragmentation works best~\cite{Nagaoka2011}. 

\section{Methods}
The experiments were performed with a COLTRIMS-setup (Cold Target Recoil Ion Momentum Spectroscopy,~\cite{doerner2000, ullrich2003}). Molecules from a supersonic gas jet are crossed orthogonal with ionizing radiation; the resulting fragments (cations and electrons) are projected by an electrostatic field onto time and position sensitive detectors. From the respective times-of-flight and the impact positions on the detector, the momentum vectors of all fragments can be determined in coincidence. As all position and time-of-flight information is stored event by event, possible correlations between fragments can be explored in the offline analysis.\\
For the results presented here, X-ray photons were used, provided by the beamline SEXTANTS at the synchrotron SOLEIL (Gif-sur-Yvette, France)~\cite{Sacchi2013}. The synchrotron was used in timing mode in order to obtain a starting time for the time-of-flight measurements. Several measurements with photon energies between 286~eV and 305~eV were performed, with energy resolutions ranging from 80~meV to 300~meV depending on the chosen size of the monochromator exit slits.\\
The molecular jet was created by expanding halothane in the gas phase through a 200~$\mu$m nozzle. The pressure before the nozzle was regulated by a needle valve to a value of 50~mbar which is significantly lower than the vapour pressure of halothane at room temperature (300~mbar). The sample was purchased from Sigma-Aldrich as a racemate (a mixture containing equal amounts of the two enantiomers) and used without further treatment. Before entering the interaction zone, the jet was collimated by two skimmers (0.3~mm diameter each). A gas recycling system with cold traps in the vacuum foreline was used to reduce sample consumption.\\
In most COLTRIMS-experiments, the electrostatic field in the spectrometer is very low (a few V/cm) in order to obtain good momentum resolution. Additionally, the ion arm of the COLTRIMS analyzer is typically very short in order to observe all fragments from a Coulomb explosion with 4$\pi$ solid angle~\cite{Jahnke2004}. This, however, entails a large overlap in the time-of-flight regions of different ion masses. When all fragment ions are detected in coincidence, the correct masses can nevertheless be assigned by checking if the calculated momenta of all the fragments add up to zero.\\
The more complex a molecule is, the higher the probability that the fragmentation produces neutral fragments which our detector is blind to. This implies that the total momentum of the measured ions does not add up to zero and thus prevents the unambiguous identification of the charged fragments as well. This is especially relevant for the case of missing hydrogen atoms.\\
To overcome this problem, a new spectrometer was designed using SIMION to simulate the electron and ion trajectories. On the ion side, a higher electric field and a long homogeneous field region, together with an electrostatic lense, proved to best fulfill the requirements. An additional field-free drift region was added in front of the ion detector, resembling a spectrometer design which is (employing low electric fields) typically used for investigation of atomic targets. This field design led to a significant improvement in the fragment identification of methyloxirane~\cite{Tia2016}.\\
An additional benefit of this design is the fact that the ions are accelerated to an energy of around 2000~eV, at which the efficiency of the microchannel-plate detectors already reaches reasonable values. This allows to omit the electrostatic meshes that are usually added in front of the detector to provide a short post-acceleration region with high electric field. As the typical transmission of the employed meshes is 0.78, they lead to a reduction in efficiency by a factor of 2.7 when attempting to measure four particles in coincidence.\\
The electron side of the spectrometer consisted simply of a homogeneous field. The high field that was determined by the ion side of the spectrometer enabled a $4\pi$ solid angle for electrons up to a kinetic energy of 35~eV. For this reason, no magnetic field was needed to confine the electron trajectories within the spectrometer. One drawback of this configuration is the reduced time-of-flight spread of the electrons, and in conjunction a poor momentum resolution along the spectrometer axis. For low-energy electrons, the good momentum resolution in the detector plane ($\Delta p \approx 0.01$ atomic units) still allows the precise determination of the electron energy and the investigation of forward-backward asymmetries in the detector plane, i.e. along the photon propagation direction. High-energy electrons, however, have a smaller acceptance angle and are mostly recorded with momenta along the spectrometer axis where the uncertainty is around 0.4 atomic units of momentum. In the case of 80~eV electrons, this leads to an experimental broadening of about 25~eV.

\section{Results and discussion}
\subsection{Determination of configuration and distinction of enantiomers}
Similar to the previous experiments, the focus was laid first on fragmentation pathways without neutral dissociation products. In this case, at least four ions need to be detected to distinguish enantiomers and to determine the absolute configuration. For halothane in the given photon energy range, the break-up into the four ions CH${}^{+}$, Cl${}^{+}$, Br${}^{+}$, and CF${}_{3}^{+}$ was the only one that fulfilled all of the above criteria. This fragmentation pathway corresponds to one of the pathways identified for CHBrClF~\cite{Pitzer2016}, with a CF${}_{3}^{+}$ fragment instead of a single F${}^{+}$ ion. The break-up into the five ions H${}^{+}$, C${}^{+}$, Cl${}^{+}$, Br${}^{+}$, and CF${}_{3}^{+}$ was found with such small yield (around 100 events per isotopic combination) that the separation of peaks for the enantiomers could not be considered statistically significant.\\
\begin{figure}
\includegraphics[width = 0.8\textwidth, keepaspectratio=true]{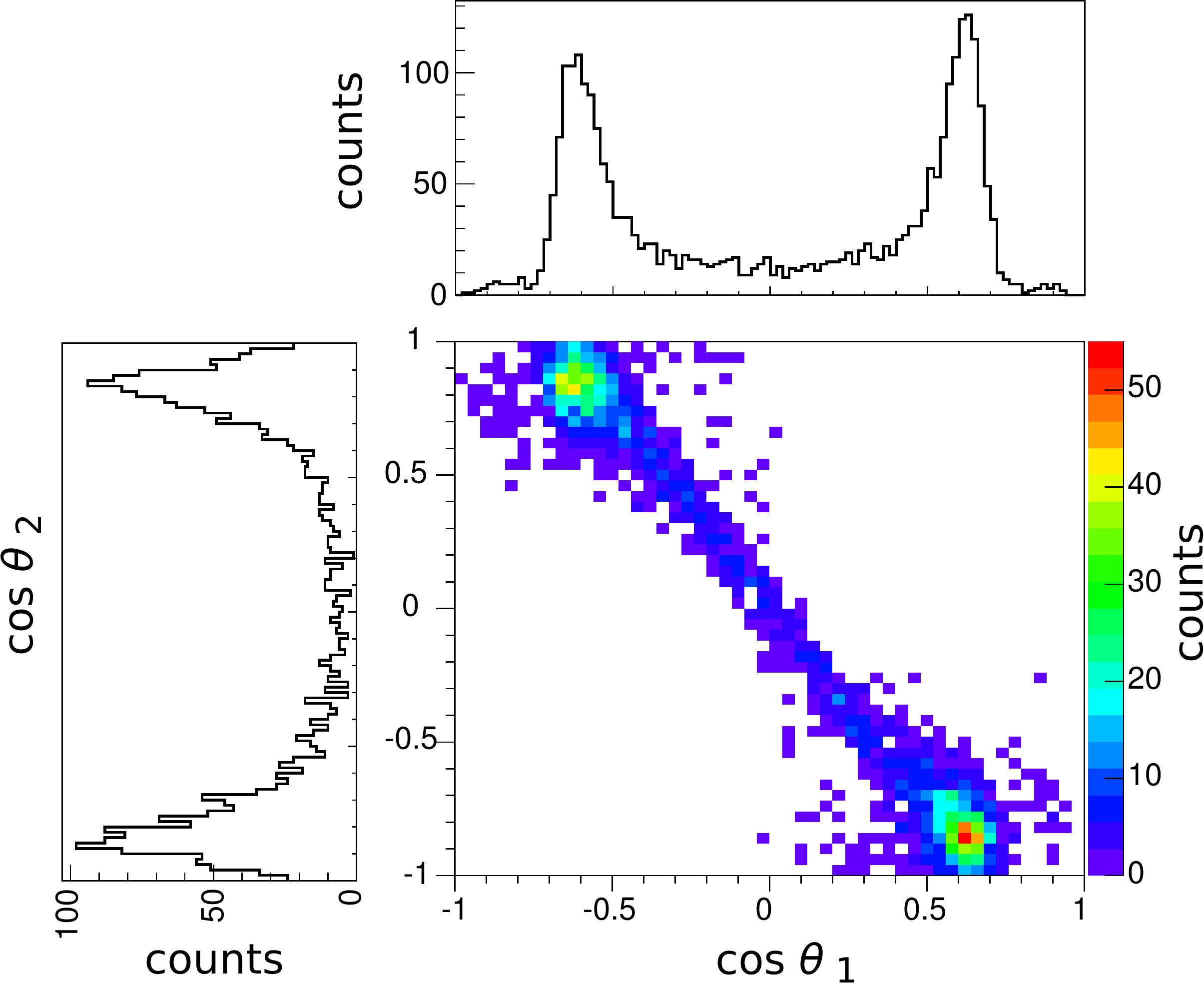}
\caption{Separation of enantiomers using the chirality parameters $\cos(\theta_1)$ and $\cos(\theta_2)$ (see text for definition) for the break-up into \{CH${}^{+}$, ${}^{35}$Cl${}^{+}$, ${}^{79}$Br${}^{+}$, CF${}_{3}^{+}$\}, obtained at a photon energy of 305~eV. The two peaks indicate a good separation of enantiomers, the location on the negative diagonal shows the consistency of the assignment. The peak in the second quadrant (top left) corresponds to the S-enantiomer, the one in the fourth quadrant (bottom right) to the R-enantiomer.}
\label{fig:halothan_305eV_chirality_parameter}
\end{figure}
Figure \ref{fig:halothan_305eV_chirality_parameter} shows the chirality parameters $\cos(\theta_1) =
\vec{p}_\mathrm{CF_3}\cdot \left(\vec{p}_\mathrm{Cl}\times \vec{p}_\mathrm{Br}\right) (|\vec{p}_\mathrm{CF_3}||\vec{p}_\mathrm{Cl} \times\vec{p}_\mathrm{Br}|)^{-1}$ $\equiv \mathrm{CF_3} \cdot (\mathrm{Cl}\times\mathrm{Br})$ and $\cos(\theta_2) =\vec{p}_\mathrm{CH}\cdot \left(\vec{p}_\mathrm{Cl}\times \vec{p}_\mathrm{Br}\right) (|\vec{p}_\mathrm{CH}||\vec{p}_\mathrm{Cl} \times\vec{p}_\mathrm{Br}|)^{-1}$ $ \equiv \mathrm{CH} \cdot(\mathrm{Cl}\times\mathrm{Br})$ plotted against each other and the one-dimensional projections. A value of $\cos(\theta_1) > 0$ corresponds to an $R$-type configuration of momenta, $\cos(\theta_1) < 0$ to an $S$-type configuration. Despite a significant number of events in between, the two peaks for the enantiomers can clearly be separated. The origin of the events close to $\cos(\theta_i) = 0$ remains unclear. One possible explanation is a fragmentation behaviour that differs from the supposed 'instantaneous' Coulomb repulsion of the ions. Another reason could be random coincidences or a falsely assigned masses, leading to erroneous momentum values. Especially events where a single C${}^{+}$ instead of a CH${}^{+}$ is detected, could contaminate the depicted distributions.\\
\begin{figure}
\includegraphics[width = \textwidth, keepaspectratio=true]{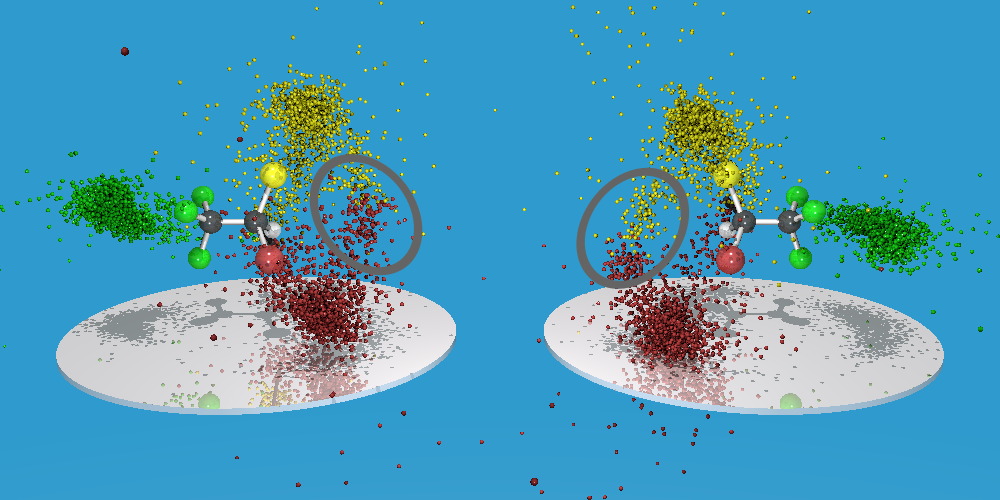}
\caption{Three-dimensional representation of the data from Figure \ref{fig:halothan_305eV_chirality_parameter}, overlaid with a structural model of the $(S)$-enantiomer (left) and $(R)$-enantiomer (right). The colour code for the model/the data is as follows: green for F/CF${}_{3}^{+}$, yellow for Cl/Cl${}^{+}$, red for Br/Br${}^{+}$, black for C/CH${}^{+}$ and white for H. The momenta of CH${}^{+}$ constitute one axis of the molecular coordinate system, and are represented by an arrow pointing away from the spectator. Ambiguous events are marked with an ellipse.}
\label{fig:halothan_305eV_3D_momenta}
\end{figure}
The three-dimensional representation (Figure \ref{fig:halothan_305eV_3D_momenta}) confirms the separation of enantiomers, with the ambiguous events clearly visible (marked by the gray ellipse). In this figure, momentum data (small spheres) are overlaid with a strucural model of the R- and S-enantiomer respectively. The transformation of the data into the molecular frame is chosen in a way that the CH${}^{+}$ points away from the spectator, defining the first axis of the molecular coordinate systems. The second axis of the molecular coordinate system is defined by the sum momentum of Cl${}^{+}$ and Br${}^{+}$.\\
\begin{figure}
\includegraphics[width = \textwidth, keepaspectratio=true]{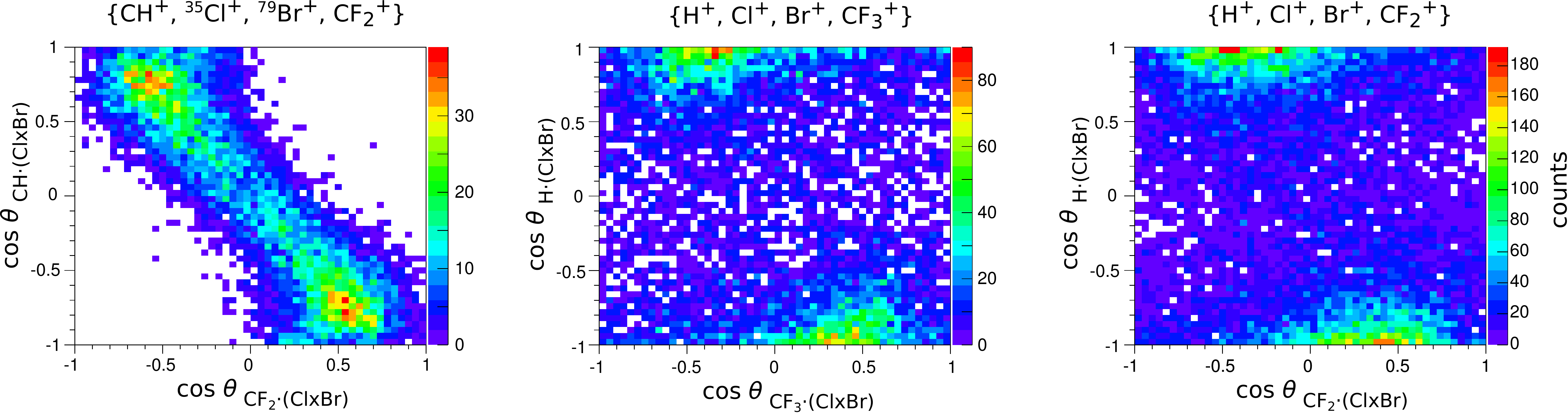}
\caption{Separation of enantiomers for additional fragmentation pathways, using the angles between ion momenta analoguous to Figure \ref{fig:halothan_305eV_chirality_parameter}. The ionizing photon energy was 305~eV as well. The break-up into \{CH${}^{+}$, ${}^{35}$Cl${}^{+}$, ${}^{79}$Br${}^{+}$, CF${}_{2}^{+}$\} (left), \{H${}^{+}$, Cl${}^{+}$, Br${}^{+}$, CF${}_{3}^{+}$\} (middle), and \{H${}^{+}$, Cl${}^{+}$, Br${}^{+}$, CF${}_{2}^{+}$\} (right) can provide a distinction between enantiomers. For the latter two break-ups, the chlorine and bromine isotopes could not be identified (see text).}
\label{fig:angles_comparison}
\end{figure}
Information on the absolute configuration can also be obtained from fragmentation pathways involving neutral dissociation products, as long as suitable ionic fragments are detected~\cite{Pitzer2016}. For these 'incomplete' break-ups, the assignment of masses becomes ambiguous: First, the isotopes of chlorine and bromine can no longer be separated; second, a contribution from C${}^{+}$ ions contaminates the CH${}^{+}$ data. Due to the high field used in the setup, the wrong mass assignment leads to errors of around 100 atomic units (a.~u.) in the momenta of the measured ions which severely affects the distinction of enantiomers. In the case of \{CH${}^{+}$, Cl${}^{+}$, Br${}^{+}$, CF${}_{2}^{+}$\}, the missing fluorine atom carries only a small momentum, leading to an average sum momentum of the detected ions of around 50 atomic units. % as can be seen in the still rather narrow line in the coincidence spectrum. 
This still allows us to apply a rather narrow constraint on the sum momentum. Thus, the contribution of the different isotopes can be separated; a clear distinction between enantiomers, however, was only visible when a restriction on the momenta of the individual ions was imposed. The maximum momentum observed in the completely detected break-up (300 atomic units) was taken as upper threshold for the momenta of the individual ions.\\
Figure \ref{fig:angles_comparison} (left) shows the resulting histogram for the chirality parameters as defined before (with $\mathrm{CF}_3 \cdot \left(\mathrm{Cl} \times \mathrm{Br}\right)$ being the shorthand notation for $\vec{p}_\mathrm{CF_3}\cdot \left(\vec{p}_\mathrm{Cl}\times \vec{p}_\mathrm{Br}\right) (|\vec{p}_\mathrm{CF_3}||\vec{p}_\mathrm{Cl} \times\vec{p}_\mathrm{Br}|)^{-1}$ etc.). Two peaks for the enantiomers, similar to Figure \ref{fig:halothan_305eV_chirality_parameter}, are visible, with an increased background in between.\\
The same condition for the ions' momenta was set for investigating the break-ups \{H${}^{+}$, Cl${}^{+}$, Br${}^{+}$, CF${}_{3}^{+}$\} and \{H${}^{+}$, Cl${}^{+}$, Br${}^{+}$, CF${}_{2}^{+}$\}, i.e. two additional Coulomb explosion pathways that are expected to yield information on the absolute configuration. In this case, a separation of the chlorine and bromine isotopes was not possible anymore. To obtain the momenta from the measured data, a mass of 35~amu was assumed for chlorine (as this isotope has a natural occurence of around 76~\%) and a mass of 80~amu for bromine (as ${}^{79}$Br and ${}^{81}$Br occur in almost equal amounts). Again two peaks are visible, but at slightly different positions than before: along the $x$-axis, the peaks are close to zero, indicating that the CF${}_{n}^{+}$ fragment almost lies in a plane with the chlorine and bromine ion. The proton H${}^{+}$ is ejected mostly perpendicular to this plane.\\
The procedure of applying constraints on the individual momenta should be employed with care as it might eliminate not only wrong mass assignments but also possible break-ups with different dynamics. In future experiments on isotope-containing species, the spectrometer design should be adjusted according to these findings. 

\subsection{Site-selective excitation}
\begin{figure}
\includegraphics[width = \textwidth, keepaspectratio=true]{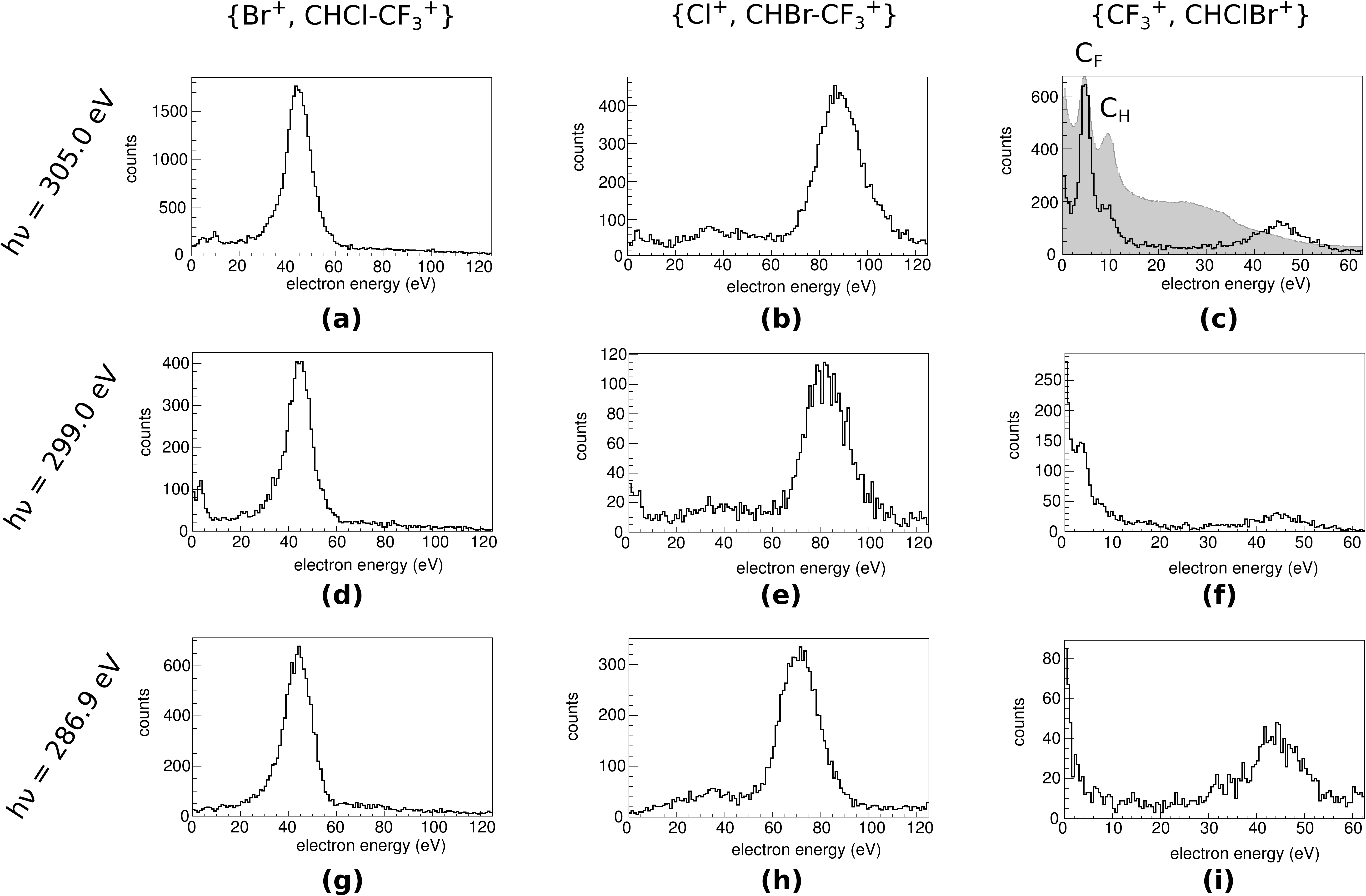}
\caption{Electron energy spectra for three different fragmentation pathways (columns) and different photon energies (rows). The constant electron energy in the left column indicates that an Auger process is dominant for the split-off of a bromine cation. In the case of the chlorine cation (center column), the shift reveals a photoelectron that can be attributed to the chlorine 2p shell. In the case of the rather symmetric break-up with a CF${}_{3}^{+}$ detached (right column), the low-energy photoelectron indicates excitation of the carbon 1s shell. Comparison with an integral electron spectrum (shaded in figure c) indicates that most ionization is from C${}_{\mathrm{F}}$. A different energy range and a different binning were chosen for the right column.}
\label{fig:halothan_2p_electron}
\end{figure}

Most chiral molecules contain several carbon atoms. This raises the question if selective excitation at the stereocenter can be used to preferentially break its bonds and to yield fragments that allow determination of absolute configuration. In halothane, the inner shells of the carbon in the CF${}_{3}$ group (we will denote this carbon as C${}_{\mathrm{F}}$ in the following) are expected to have a larger binding energy than the corresponding levels of the stereocenter C${}^{\star}$. This is due to the higher electronegativity of fluorine compared to bromine, chlorine and hydrogen.\\ % due to the strongly electronegative fluorine atoms. 
A scan over a photon energy range from 280~eV to 305~eV was performed, recording the total ion yield and the yield of CF${}^{+}_{3}$ ions. As the features could not unambiguously be assigned to resonant excitations of one of the carbon atoms, the following photon energies were chosen: 286.9~eV (below excitation of the carbon atoms), 299.0~eV (photoionization of the stereocenter only) and 305.0~eV (photoionization of both carbon atoms). For the latter two energies, the coincident measurement of the electron energy allows us to determine the excitation site. The first one was chosen to evaluate the influence of excitation from different halogen shells that have previously been found to play a role in the multiple ionization~\cite{Pitzer2016}.\\
Confirming our previous work~\cite{Pitzer2016}, the electron energy spectra for the fragmentation into four cations do not show any structures. This could be due to the physical processes leading to this fragmentation or due to experimental limitations that do not allow to detect four electrons in coincidence correctly. An effect of site-specific excitation, if any, can thus only be found in the relative abundance of the different fragmentation pathways. Table \ref{tab:yields} shows the yield of the four fragmentation pathways discussed in the previous section, normalized to the total ion yield (measured via the total number of events).
\begin{table}%
	\begin{tabular}[h!t]{|l|l|l|l|}
	\hline
detected fragments & \multicolumn{3}{|c|}{photon energy} \\ \hline	
 & 305.0 eV & 299.0 eV & 286.9 eV\\ \hline 
\{CH${}^{+}$, CF${}_{3}^{+}$, Cl${}^{+}$, Br${}^{+}$\} & $3.9 \cdot 10^{-5}$ & $2.7 \cdot 10^{-5}$ & $4.7 \cdot 10^{-5}$  \\
\{CH${}^{+}$, CF${}_{2}^{+}$, Cl${}^{+}$, Br${}^{+}$\} & $1.8 \cdot 10^{-4}$ & $1.2 \cdot 10^{-4}$ & $2.1 \cdot 10^{-4}$ \\  
\{H${}^{+}$, CF${}_{3}^{+}$, Cl${}^{+}$, Br${}^{+}$\} & $\mathbf{1.2 \cdot 10^{-4}}$ & $3.4 \cdot 10^{-5}$ & $3.2 \cdot 10^{-5}$ \\  
\{H${}^{+}$, CF${}_{2}^{+}$, Cl${}^{+}$, Br${}^{+}$\} & $\mathbf{2.8 \cdot 10^{-4}}$ & $2.9 \cdot 10^{-5}$ & $6.4 \cdot 10^{-5}$ \\  
 \hline %\hline \noalign{\smallskip}
\end{tabular}%
\caption{Yield of different four-body fragmentation pathways normalized to the total number of events. The relative yield of the first two pathways does not change when going above the carbon 1s ionization threshold. The fragmentation pathways with a detected proton are slightly enhanced when C${}_{\mathrm{F}}$ can be photoionized (bold font).}
%from: yields.ods
%1. Zeile: impulsfenster (20,20,20) um die vier Isotopenlinien (in der jeweils korrekten Zuordnung: .root) und addiert
%2. Zeile: impulsfenster (60,60,60) um die vier Isotopenlinien (in der jeweils korrekten Zuordnung: .root) und addiert
%alle Zeilen: ion p < 300 a.u.
\label{tab:yields}
\end{table}
The first two break-ups, involving a CH${}^{+}$ ion do not show significant dependence of the photon energy. The break-ups where a proton was detected seem to have a higher yield by a factor of 3 when the photon energy is above the ionization threshold of C${}_{\mathrm{F}}$. As the yield of the \{H${}^{+}$, CF${}_{2}^{+}$, Cl${}^{+}$, Br${}^{+}$\} does not show a clear tendency, the numbers shown do not provide a watertight indication of selective excitation. In addition, an increase is expected above the photoionization threshold of the stereocenter (i.e. at 299~eV in our experiment), as the hydrogen is bound to C${}^{\star}$.\\ 
We therefore chose to additionally investigate the selective excitation for fragmentation into two cations where clear features in the electron spectra can be seen. Three break-ups could be identified that are 'complete', i.e. where the sum of fragment masses corresponds to the parent ion's mass: \{Br${}^{+}$, CHClCF${}_{3}^{+}$\} (\textbf{I}), \{Cl${}^{+}$, CHBrCF${}_{3}^{+}$\} (\textbf{II}), and \{CF${}_{3}^{+}$, CHClBr${}^{+}$\} (\textbf{III}).\\
The electron spectra differ significantly for the three break-ups, indicating that they are induced by different electron dynamics. Figure \ref{fig:halothan_2p_electron} shows the energy spectra for different photon energies (rows) and fragmentation pathways (columns). 
Pathway \textbf{I}: When a bromine ion is split from the molecule, the corresponding electron spectrum shows a peak at around 45~eV. As this peak does not shift with the excitation energy, it is attributed to an Auger decay. The peak position and width correspond to the electron energy distribution discussed in Ref.~\cite{Miron2009} and attributed there to valence-valence Auger processes after the bromine 3d excitation. Surprisingly, no photoelectron was found in this break-up channel for any of the exciting energies. This could be due to the fact that the solid angle for electrons with an energy above 200~eV (the bromine 3d level being at 81~eV) was very small. Additionally, an electron of a kinetic energy of 200 eV is detected with a resolution of 45~eV and thus dramatically smeared out in the spectrum. \\
Pathway \textbf{II}: In the case of a separated chlorine ion, a peak is observed in the spectrum that shifts roughly with the photon energy. In this case, the experimental broadening discussed in the experimental section plays a significant role. Although a precise determination of the energy is not possible, the attribution as a photoelectron from the chlorine 2p level (206.8 eV for the atomic species~\cite{Yeh1985}) seems justified.\\
Pathway \textbf{III}: Only the break-up involving CF${}_{3}^{+}$ is clearly connected to the excitation of the C 1s shell: For the highest photon energy, a distinct peak is observed that corresponds to the lower electron energy in the integral electron spectrum (shaded area), together with a shoulder at the higher electron energy. This indicates that the break-up is preferentially induced by excitation of the 1s shell of C${}_{\mathrm{F}}$. The spectra at $h\nu = 299.0$~eV and $h\nu = 286.9$~eV support this finding. In the former one, the strong peak has disappeared, in the latter one, no sign of a photoelectron is visible. The relatively high yield at 299.0~eV might be due to the fact that the energy is very close to the threshold of C${}_{\mathrm{F}}$. 
These results show that a specific fragmentation pathway can strongly correlate to the selective excitation of one carbon atom.\\
In the case of halothane, various shells from the halogen atoms contribute to the ionization, thus preventing a truly selective excitation. At the photon energies used, the cross sections for photoionization from the chlorine 2p and the bromine 3d level are even higher than for the targeted carbon 1s. For oxygen and nitrogen, in contrast, the photoionization cross section at 300~eV is more than an order of magnitude lower than for carbon~\cite{Yeh1985}. The probability for alternative ionization pathways is thus expected to decrease significantly for organic species that contain only oxygen or nitrogen in addition to carbon and hydrogen. The crucial question in this case will be if the levels of the carbon atoms differ enough to enable a selective excitation.

\section{Conclusion}
The chiral ethane derivative halothane was investigated using X-ray photons for core ionization and a COLTRIMS-setup for coincident detection of the fragments. A separated signal of the two enantiomers could be found in several four-body Coulomb explosion pathways. The CF${}_{3}^{+}$ cation is an example of a complete functional group that was used to determine absolute configuration. This finding is an encouraging step towards application of the method to biologically relevant molecules. Due to multiple possible fragmentation pathways, however, the yield of chirality-sensitive pathways was very low.\\
The site-selective excitation of the two carbon atoms was investigated for four-body fragmentations and in the case of double ionization, leading to two-body fragmentation. Two of the four-body fragmentation pathways show an increase for photon energies where both carbon atoms can be photoionized.
In the case of two-particle break-ups, the splitting of the CF${}_{3}^{+}$ from the rest of the molecule clearly correlates with excitation of the C${}_{\mathrm{F}}$. For other break-ups, the contribution of chlorine and bromine shells is dominant, providing rather element-selectivity than a site-selectivity.\\
The results presented here show that coincidence measurements can yield information on the absolute configuration of molecules containing more than one carbon atom. These results are a further step towards analytical applications of the method and provide a good example for linking site-selective excitation with investigations of molecular structure.

\section{Acknowledgments}
We thank the staff of the synchrotron SOLEIL, in particular Nicolas Jaouen from
beamline SEXTANTS for their outstanding support. Markus Sch{\"o}ffler acknowledges support by the Adolf Messer foundation.
This work was supported by the State Initiative for the 
Development of Scientific and Economic Excellence (LOEWE) in the 
LOEWE-Focus ELCH. 
\bibliographystyle{unsrt}  
%\bibliography{halothan}

\end{document}